\documentclass[prl, showpacs, twocolumn, floatfix]{revtex4}

\usepackage{graphicx}
\usepackage{amsmath, amsfonts, amssymb, bm}
\usepackage[]{psfrag}
\usepackage [dvips]{epsfig}
\usepackage [hang, tight]{subfigure}
\usepackage {dsfont}
\usepackage {color}

\psfragscanoff
\begin{document}

\title{Pair production in laser fields oscillating in space and time}
\author{Matthias \surname{Ruf}}
\author{Guido R. \surname{Mocken}}
\author{Carsten \surname{M\"{u}ller}}
\email{carsten.mueller@mpi-hd.mpg.de}
\author{Karen Z. \surname{Hatsagortsyan}}
\author{Christoph H. \surname{Keitel}}
\affiliation{Max-Planck-Institut f\"ur Kernphysik, Saupfercheckweg 1, D-69117 Heidelberg, Germany}

\date{\today}

\begin{abstract}
The production of electron-positron pairs from vacuum by counterpropagating laser beams of linear polarization is calculated. In contrast to the usual approximate approach, the spatial dependence and magnetic component of the laser field are taken into account. We show that the latter strongly affects the creation process at high laser frequency: the production probability is reduced, the kinematics is fundamentally modified, the resonant Rabi-oscillation pattern is distorted and the resonance positions are shifted, multiplied and split.
\end{abstract}


\maketitle
In the presence of very strong electromagnetic fields the quantum electrodynamic vacuum becomes unstable and decays into electron-positron ($e^+e^-$) pairs \cite{Schwinger}. 
This phenomenon has been realized experimentally, e.g., in relativistic heavy-ion collisions \cite{HeavyIonCol}. Shortly after the invention of the laser almost 50 years ago, theoreticians began to study $e^+e^-$ pair production by intense laser light \cite{Itzykson}. Because of the remarkable progress in laser technology during recent years, the interest in the process has been revived since an experimental investigation of pair creation by pure laser light is coming into reach. The only observation of laser-induced pair production until now was accomplished ten years ago at SLAC (Stanford, California), where a 46 GeV electron beam was brought into collision with an intense optical laser pulse \cite{SLAC}. 
In this experiment, a $\gamma$-photon produced via Compton scattering or the electron Coulomb field assisted the laser beam in the pair production.

The most simple field configuration for realization of purely laser-induced pair production consists of two counterpropagating laser pulses of equal frequency and intensity (see \cite{Rabi_Osc,PP_XFEL1,PP_XFEL2,finite_size} and references therein). The resulting standing wave is inhomogeneous both in time and in space which represents a formidable task for the nonperturbative quantum field theory, see e.g. \cite{GiesDunne}. All theoretical investigations so far have approximated the standing laser wave by a spatially homogeneous electric field oscillating in time. This dipole approximation is expected to be well-justified in optical laser fields, where the wavelength is much larger than the typical length scale of the process: $\lambda\gg l\sim 2m/(|e|E)$ in natural units $(\hbar=c=1)$ which are employed throughout.  In terms of the relativistic parameter $\xi=|e|E/(m\omega)$, this relation corresponds to $\xi\gg 1$. Here $E$ and $\omega$ are the laser field and its frequency, $e$ and $m$  the electron charge and mass, respectively. 
Meanwhile the experimental realization of laser-induced pair production is also extensively discussed in connection with upcoming x-ray free-electron laser (XFEL) facilities \cite{PP_XFEL1,PP_XFEL2}. In this case, however, the laser frequency is high, $\xi\lesssim 1$ and the magnetic field component is not negligible. The latter, in general, can have an important influence on the pair creation process. This is most evidently demonstrated by the fact that a single plane laser wave cannot extract pairs from the vacuum, whereas a purely electric field can. 

In this Letter, we calculate $e^+e^-$ pair creation by two counterpropagating strong laser pulses (CLP) of high frequency ($\omega \lesssim m$ and $\xi\lesssim 1$), taking into account explicitly the temporal and the spatial dependence of the laser fields and their magnetic components. The pair production is modelled via an electron transition from an initial negative-energy state to a final positive one, based on a solution of the Dirac equation \cite{Fradkin}. By numerically solving the latter, we demonstrate that the laser magnetic field component significantly reduces the pair creation rate. Moreover, it leads to distinct qualitative changes in the process as the resonant Rabi oscillation pattern is strongly disturbed. These features can be used to test strong-field QED in temporally and spatially inhomogeneous fields.

\begin{figure}[b]
   \centering\includegraphics[clip,height=4.0cm]{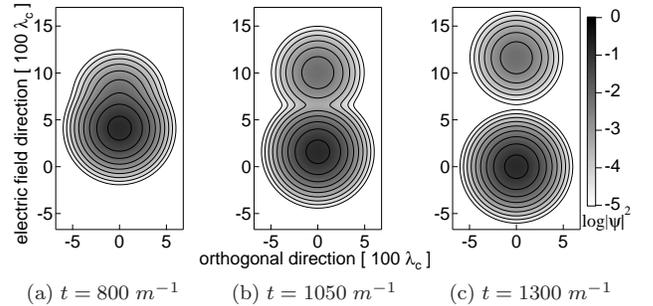}\hfill
   \caption[]{Pair creation in an oscillating electric field with frequency $\omega=m/200$ and 
             critical field strength $E_c=m^2/e$. Three snapshots of the probability distribution are taken at times 
             as indicated from the evolution of an initially negative-energy Gaussian wave-packet at rest at the origin.
             The length is scaled in multiples of the Compton wavelength $\lambda_c=1/m$.}
    \label{fig:baby}
\end{figure}

We employ an advanced computer code that solves the Dirac equation in an arbitrary external potential on a two-dimensional spatial grid \cite{Guido}. An initial wave-packet $\Psi(t=0)$ in the negative-energy continuum, representing an electron in the Dirac sea, is propagated via the split-operator algorithm. Under the influence of the external field an $e^+e^-$ pair can be produced. The transition amplitude is  determined by projection of the wave function onto positive-energy states $\Phi_{{\bf p}^{\prime}}^{(+)}$ after the external field has been turned off. In this approach an intuitive graphical interpretation for the production process is possible. As an example Fig.~\ref{fig:baby} shows the time evolution of an initially negative-energy Gaussian wave packet at rest under the influence of an oscillating electric field (OEF). When the $e^+e^-$ pair is produced, a droplet is separated from the wave packet which moves opposite to the initial one. The droplet is a positive energy state and represents the created electron. The change of the sign of energy is evidenced by the change of the group velocity of the droplet wave packet. 

Our numerical approach enables us to take into account the full space-time dependence of the laser potential and to investigate the effect of the magnetic field. Before elaborating on the differences  of the pair production dynamics in an OEF and CLP, we briefly summarize the main results for pair production in an OEF.
\begin{figure}
    \centering\includegraphics[clip,width=0.85\linewidth]{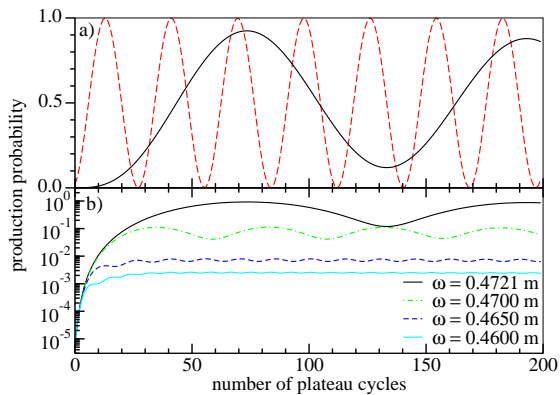}\hfill
    \caption[]{(color online) Disturbed Rabi oscillations: Pair production probability versus pulse length 
               for fixed turn-on and turn-off phases of $\sin^2$-shape and a duration of half a cycle each at $\xi=1$.
               (a) The red dashed line and the black solid line show the OEF case for $\omega=0.49072 m$
               and the CLP case for $\omega=0.4721 m$, respectively, each corresponding to a $5$-photon 
               resonance. (b) CLP case for various frequencies. The black solid curve coincides with the resonant 
               case in part (a), while the others correspond to off-resonant situations.} 
    \label{fig:rabi}
\end{figure}
\begin{figure}
    \centering\includegraphics[clip,width=0.85\linewidth]{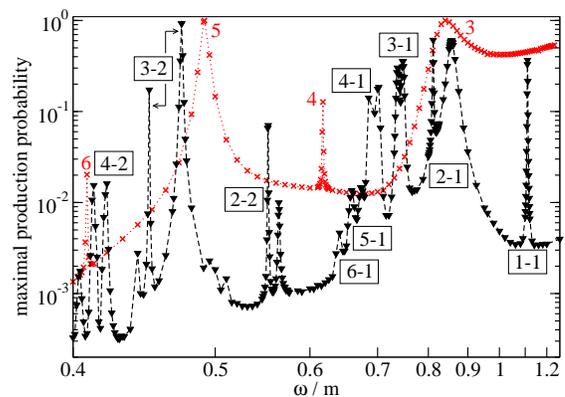}\hfill
    \caption[]{(color online) Resonant probability spectrum: Maximal value of the pair production probability during Rabi 
               oscillation at $\xi=1$, varying the pulse length up to 200 cycles. The red crosses show the OEF spectrum; 
               the peak labels denote the absorbed photon number. The black triangles show the CLP spectrum. 
               Here the labeling signifies the number of absorbed photons from the right-left propagating 
               waves. A splitting occurs, as indicated by arrows for the example of the ($3$-$2$) peak. The frequency axis is plotted 
               reciprocally \cite{num}.}
    \label{fig:resonance}
\end{figure}
Due to momentum conservation in OEF, the problem is reduced to a two-level system consisting of a negative and a positive-energy state where a multiphoton resonance $2q_0({\bf p})=n\omega$ enforced by energy conservation can exist, leading to Rabi oscillations between these states \cite{Rabi_Osc}. Here $q_0({\bf p})=(1/T)\int_0^T\mathrm{d}t\sqrt{\left({\bf p}-e{\bf A}(t)\right)^2+m^2}$ is the quasi-energy of the laser-dressed state, ${\bf p}$ the canonical momentum, $T$ the pulse duration and $n$ the number of absorbed photons. The relativistic parameter $\xi$ distinguishes between different interaction regimes of pair creation in laser fields. For $\xi\ll 1$ the process probability follows a perturbative power law dependence: $W \sim \xi^{2n}$, whereas for $\xi\gg 1$ ($E<E_c$) a tunneling behavior: $W \sim \exp(-\pi E_c/E)$, where $E_c=m^2/|e|$ is the critical field \cite{Itzykson}. Most interesting is the nonperturbative regime ($\xi\lesssim 1$), where non-dipole dynamics is pronounced and no simple asymptotic formulas exist. Therefore, we restrict the following discussion to $\xi=1$ (for both laser beams taken together).
\begin{figure*}[t]
    \centering\includegraphics[clip,width=0.8\linewidth]{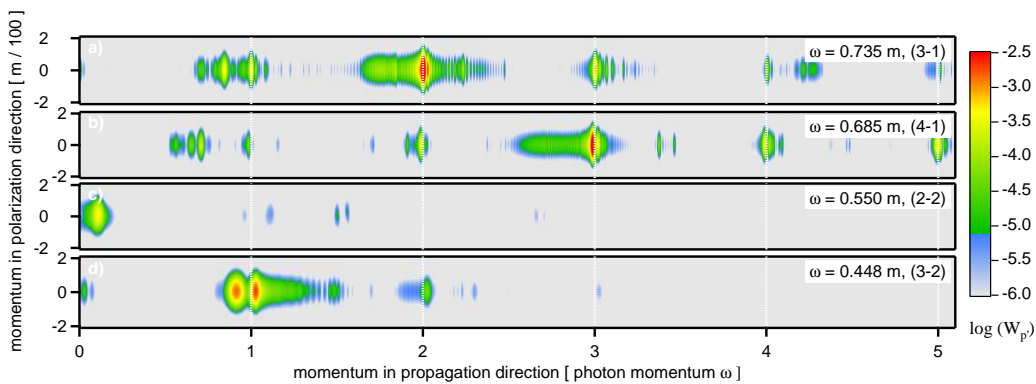}\hfill
    \caption[]{(color online) Final positive-energy momentum distributions 
               $W_{p^{\prime}}=|\!\!<\!\Phi_{{\bf p}^{\prime}}^{(+)}|\Psi(T)\!>\!\!|^2$ after the interaction 
               with two counterpropagating laser pulses ($\xi=1$, $T=150\pi/\omega$). Due to its magnetic component 
               the field transfers longitudinal momentum to the wave-packet. Shown are the results for four different 
               frequencies corresponding to four peaks in the resonance spectrum in Fig.~\ref{fig:resonance}. Due 
               to the symmetry of the spectra under momentum inversion $p_z^{\prime}\rightarrow-p_z^{\prime}$, we 
               only show the positive half of them.} 
    \label{fig:spectra}
\end{figure*}

Now we turn to pair production in CLP. Our calculation always assumes an initially narrow Gaussian wave-packet centered around ${\bf p}=0$ of width $\Delta p\approx\alpha m\ll m$, with the fine structure constant $\alpha$, lying in the negative-energy continuum. A particular spin state along the magnetic field direction is chosen; the opposite spin orientation would give identical results. The wave-packet is exposed to two linearly polarized laser pulses counterpropagating along the $z$-axis and featuring $\sin^2$ turn-on and turn-off phases of half a cycle each. The number of plateau cycles with constant intensity is variable. To save computing time the fields are switched off when they do not overlap anymore. The pair production probability is expected to depend on the pulse length as known from the OEF case. The latter is shown in Fig.~\ref{fig:rabi}\,(a), where the red dashed curve shows the expected Rabi oscillation in an OEF for $\omega=0.49072m$ corresponding to an $n=5$ photon resonance, because the value of the quasi-energy is $q_0(0)\approx 1.21m$ which is independent of frequency for given $\xi$. To compare this with the CLP case one has to find the $n=5$ resonance frequency. In this case $q_0$ is no longer analytically computable. Fig.~\ref{fig:rabi}\,(b) shows the result for various frequencies starting from $\omega=0.46m$ on a logarithmic scale. The resonance frequency for $n=5$ turns out to be $\omega=0.4721m$. The identification of resonances in the CLP case will be explained below. The oscillation pattern is strongly modified by inclusion of the laser magnetic field, which is seen most clearly in Fig.~\ref{fig:rabi}\,(b). The probability oscillates around a plateau value with a frequency five times smaller than the Rabi frequency $\Omega$ in OEF. Note that the  pair production rate at small times $t\ll \Omega^{-1}$ scales as $\Omega^2$ \cite{Rabi_Osc} and is reduced due to the magnetic field effect by 1-2 orders of magnitude.

Taking for each frequency the maximum of the production probability as a function of laser pulse duration, one arrives at the spectrum in Fig.~\ref{fig:resonance}. In that way the normally superimposed pulse length-dependent oscillations of the probability are omitted, and the resonances are clearly visible. In the OEF case, only odd-$n$ resonances occur for ${\bf p}=0$, as the even ones are forbidden by a charge-conjugation related selection rule \cite{footnote:C}. Due to the non-zero wave-packet width $\Delta p$, even-$n$ resonances do occur in Fig.~\ref{fig:resonance}, but they are suppressed as compared to the odd ones. The picture changes significantly when real laser fields are applied: The height of the probability spectrum is reduced by approximately one order of magnitude, the resonances are shifted, several new resonances occur, and the resonance lines are split. 

In order to explain these modifications, we examined the corresponding momentum distributions of the created electron as shown in Fig.~\ref{fig:spectra} for the peaks labeled by ($3$-$1$), ($4$-$1$), ($2$-$2$) and ($3$-$2$). In contrast to the OEF case, the photons in the CLP carry momentum along the beam axis, which is transferred to the electron wave-packet upon absorption. Only the transversal momentum components are conserved here. By energy-momentum conservation, a number of $n_+$ ($n_-$) photons absorbed from the beam propagating to the right (left) determine the laser-dressed final particle 4-quasi-momenta:
\begin{eqnarray}
   q^{\prime}=n_+k_++n_-k_--q,
   \label{eq:energy}
\end{eqnarray}
where $q$ is the electron initial 4-quasi-momentum, $k_{\pm}=(\omega,0,0,\pm \omega)$ the photon 4-momenta for the right/left propagating waves. Our numerical calculations of the final momentum distribution after the laser fields have passed in Fig.~\ref{fig:spectra} confirm Eq.~(\ref{eq:energy}): the mainly contributing region for each peak corresponds to $p_z^{\prime}=(n_+-n_-)\omega$, for example $p_z^{\prime}\approx 2\omega$ at $\omega=0.735m$. The latter means that the  final dressed momenta $q_z^{\prime}$ do not differ essentially from the momenta outside the laser field $p_z^{\prime}$. In order to determine the resonance frequencies, we assume that the effective mass $m_*$  depends only on $\xi$ as in the OEF case ($q^2=m_*^2$) and that the initial quasi-momentum vanishes (${\bf q}=0$) because of the initial vanishing momentum ${\bf p}=0$. Then, the resonance frequencies read
\begin{equation}
   \omega=\frac{m_*}{2}\frac{n_++n_-}{n_+n_-}. 
   \label{eq:frequency}
\end{equation}
The main contribution to the momentum spectrum for the peak at $\omega=1.1m$ in Fig.~\ref{fig:resonance} comes from $p_z^{\prime}=0$, when the number of absorbed photons from the left and right laser beam are the same. This peak belongs to the lowest possible photon number $n_+=n_-=1$, resulting in $m_*=1.11m$. Then all peaks can be explained by Eq.~(\ref{eq:frequency}) and correspond to the given labeling ($n_+\,$-$\:n_-$). The interchange of $n_+\leftrightarrow n_-$ gives the same resonant frequency; we chose $n_+\ge n_-$ for the labeling. All resonances with $n_-=1$ lie above $\omega=m_*/2$, according to Eq.~(\ref{eq:frequency}). We found them up to a photon number of $n_++1=7$. Their height decreases with increasing photon number.

\begin{figure}[t]
   \centering\includegraphics[clip,width=\linewidth]{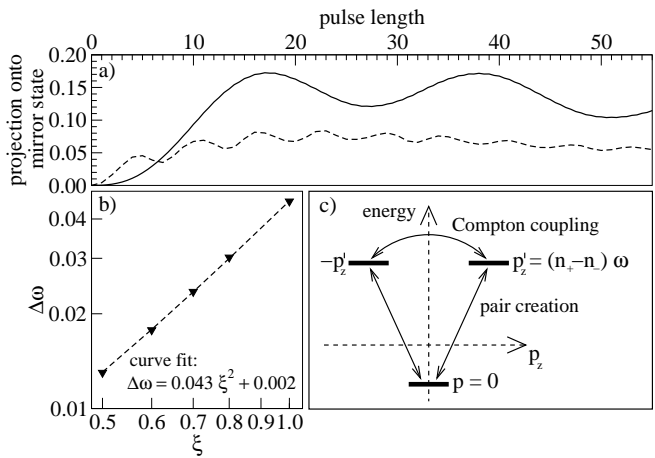}\hfill
   \caption[]{Coupling between the two upper states of the V-type system via Compton-scattering for 
              the ($2-1$) peak. a) Population of the initially empty mirror state with 
              $p_z^{\prime}=-p_{\rm initial}=-\omega$: (solid line) $\omega=0.766m$ and $\xi=0.5$, 
              resulting in the Rabi frequency of $\Omega_C=0.018m$ due to Compton scattering; 
              (dashed line) $\omega=0.812m$ and $\xi=1.0$ resulting in $\Omega_C=0.066m$. In both 
              cases $\Omega\ll\Omega_C$, where $\Omega$ is the associated Rabi frequency due to pair creation.
              b) $\xi$ dependence of the peak splitting, $\Delta\omega\sim\xi^2$. c) Level scheme.}
   \label{fig:compton}
\end{figure}

So far we have explained the overall structure of the resonance spectrum in Fig.~\ref{fig:spectra}. There is however a substructure inherent to all resonance peaks with $n_+\ne n_-$. The resonances are split into double peaks which do not occur for an OEF. The reason for this is again the spatial dependence of the laser fields  so that the photons carry momentum in propagation direction. The negative-energy electron can absorb $n_+$ photons from the left beam and $n_-$ from the right, or vice versa, reaching two different final positive energy states with equal energy but opposite momentum. Therefore, the pure two-level system of the OEF is transformed into a V-type three-level system (if $n_+\ne n_-$) in CLP, see Fig.~\ref{fig:compton}\,(c). The positive-energy states are coupled via Compton scattering as in the quantum regime of free electron lasers. 
This can lead to splitting of these levels and, consequently, to splitting of the resonant transition from the initial state, analogous to the Autler-Townes effect \cite{Autler-Townes}. We investigated the splitting for the ($2\,$-$\:1$) resonance. Fig.~\ref{fig:compton}\,(a) shows the Compton oscillation between the two positive-energy states for two different $\xi$ values. The Rabi frequency $\Omega_C$ due to Compton scattering increases by a factor of $3.6$ when $\xi$ is varied from 0.5 to 1. The splitting for various $\xi$ is shown in Fig.~\ref{fig:compton}\,(b). A quadratic dependence $\Delta\omega\sim\xi^2$ is found, leading to an increase of the splitting from $\xi=0.5$ to $\xi=1.0$ by a factor of $3.5$. This indicates that the observed peak fine structure of the spectra is indeed an Autler-Townes-like effect. An equivalent explanation of the peak splitting can also be offered:
The spatial periodicity of the field induces a band structure of the electron energies \cite{Becker} which inhibits electron creation in certain energy regions. The splitting becomes larger with increasing $\xi$, following the energy gap behavior.

Laser sources which could in principle be employed to enter the parameter range of interest are the XFEL facilities presently under construction at SLAC and DESY (Hamburg, Germany). They are envisaged to provide coherent synchrotron radiation with photon energies up to 12~keV and could reach the $\xi\approx 1$ regime if the laser beam is focused to sub-$\mu$m waist size \cite{PP_XFEL1}. Powerful hard x-ray sources are also expected via relativistic laser-plasma interaction \cite{plasma}.

In conclusion, we employed a numerical approach to investigate $e^+e^-$ pair creation in counterpropagating laser fields. In comparison to the OEF case, we find characteristic modifications of the created particle spectra and the Rabi oscillation dynamics. The narrow peak splitting of the resonant pair production probability could serve as a sensitive probe of the quasi-energy band structure and, generally, of QED in superstrong spatially and temporally inhomogeneous fields. In particular, the potential relevance of non-standard model field theories \cite{nonstandardGies} could be tested.

We acknowledge enlightening discussions with D. Bauer.


\begin{thebibliography}{9}

\bibitem{Schwinger}
F.~Sauter, Z. Phys. {\bf 69}, 742 (1931);
W.~Heisenberg and H.~Euler, {\it ibid.} {\bf 98}, 714 (1936);
J.~Schwinger, Phys. Rev. {\bf 82}, 664 (1951).

\bibitem{HeavyIonCol}
A.~Belkacem {\it et al.}, Phys. Rev. Lett. {\bf 71}, 1514 (1993). 

\bibitem{Itzykson} 
E.~Brezin and C.~Itzykson, Phys. Rev. D {\bf 2}, 1191 (1970);
V.~S.~Popov, JETP Lett. {\bf 13}, 185 (1971).

\bibitem{SLAC} 
D.~Burke {\it et al.}, Phys. Rev. Lett. {\bf 79}, 1626 (1997).

\bibitem{Rabi_Osc} 
V.~S.~Popov, JETP Lett. {\bf 18}, 255 (1973); 
N.~B.~Narozhny and A.~I.~Nikishov, Sov. Phys. JETP {\bf 38}, 427 (1974); 
V.~M.~Mostepanenko and V.~M.~Frolov, Sov. J. Nucl. Phys. {\bf 19}, 451 (1974);
H. K. Avetissian {\it et al.}, Phys. Rev. E {\bf 66}, 016502 (2002).

\bibitem{finite_size} 
W.~Becker, Laser Part. Beams {\bf 9}, 603 (1991);
V.~S.~Popov, JETP Lett. {\bf 74}, 133 (2001); 
N. B. Narozhny {\it et al.}, {\it ibid.} {\bf 80}, 382 (2004);
A.~Di~Piazza, Phys. Rev. D {\bf 70}, 053013 (2004);
P.~Krekora {\it et al.}, Phys. Rev. Lett. {\bf 95}, 070403 (2005);
D.~B.~Blaschke {\it et al.}, {\it ibid.} {\bf 96}, 140402 (2006);
S.~S.~Bulanov {\it et al.}, JETP {\bf 102}, 9 (2006);
M.~Marklund {\it et al.}, JETP Lett. {\bf 83}, 313, 2006;
M.~V.~Fedorov, M.~A.~Efremov, and P.~A.~Volkov, Opt. Comm. {\bf 264}, 413 (2006).

\bibitem{PP_XFEL1} 
A.~Ringwald, Phys. Lett. B {\bf 510}, 107 (2001).

\bibitem{PP_XFEL2}
R.~Alkofer {\it et al.}, Phys. Rev. Lett. {\bf 87}, 193902 (2001); 
C.~D.~Roberts, S.~M.~Schmidt, and D.~V.~Vinnik, {\it ibid.} {\bf 89}, 153901 (2002).

\bibitem{GiesDunne} 
H.~Gies and K.~Klingm\"uller, Phys. Rev. D {\bf 72}, 065001 (2005);
G.~V.~Dunne and C.~Schubert, {\it ibid.} {\bf 72}, 105004 (2005).

\bibitem{Fradkin} 
E.~S.~Fradkin, D.~M.~Gitman and Sh.~M.~Shvartsman, {\it Quantum Electrodynamics with unstable vacuum} (Springer, Berlin, 1991).

\bibitem{Guido} 
G.~R.~Mocken and C.~H.~Keitel, J. Comp. Phys. {\bf 199}, 558 (2004); Comput. Phys. Comm. {\bf 178}, 868 (2008).

\bibitem{num} 
We consider $\omega\ge0.4m$ in Fig.~\ref{fig:resonance} because of computational difficulties at lower frequencies.

\bibitem{footnote:C}
For zero momentum, the final $e^+e^-$ state is odd under charge-conjugation. This can only be achieved by the absorption of an odd number of photons.


\bibitem{Autler-Townes} 
S.~H.~Autler and C.~H.~Townes, Phys. Rev.  {\bf 100}, 703 (1955).

\bibitem{Becker} 
For an electron moving in a standing wave of circular polarization, this is shown in 
W.~Becker, R.~Meckbach and H.~Mitter, J. Phys. A  {\bf 12}, 799 (1979).

\bibitem{plasma}  
G.~D.~Tsakiris {\it et al.}, New J. Phys. {\bf 8}, 19 (2006);
S.~V.~Bulanov, T.~Esirkepov and T. Tajima, Phys. Rev. Lett. \textbf{91}, 085001 (2003); 
S.~Gordienko {\it et al.}, {\it ibid.} {\bf 94}, 103903 (2005).

\bibitem{nonstandardGies}
H.~Gies, J. Phys. A {\bf 41}, 164039 (2008).

\end{thebibliography}
\end{document}